\documentclass[journal]{IEEEtran}
\usepackage{multicol}
\usepackage{graphicx}
\usepackage{cleveref}
\usepackage{epstopdf}
\usepackage{bm}
\begin{document}

\title{Intermodal Four-Wave-Mixing and Parametric Amplification in km-long Fibers}

\author{Massimiliano~Guasoni,
        Francesca~Parmigiani, Peter~Horak, Julien~Fatome,
        and~David~J.~Richardson% <-this % stops a space
\thanks{Manuscript received XXX.}% 
\thanks{M. Guasoni is supported through an Individual Marie Sklodowska-
Curie Fellowship (H2020 – MSCA – IF – 2015 , project AMUSIC - Grant Agreement 702702). This research is sponsored by EPSRC grant EP/P026575/1.}%
\thanks{M. Guasoni, F. Parmigiani, P. Horak, and D. J. Richardson are with the Optoelectronics Reserch Centre, University of Southampton, Southampton SO17 1BJ, United Kingdom.}%
\thanks{J. Fatome is with the Laboratoire Interdisciplinaire Carnot de Bourgogne (ICB), UMR 6303 CNRS - Universit\'e Bourgogne Franche-Comt\'e,
9 Avenue Alain Savary, BP 47870, 21078 Dijon, France}%
\thanks{M. Guasoni's email is m.guasoni@soton.ac.uk}}


\maketitle

\begin{abstract}
We theoretically and numerically investigate intermodal four-wave-mixing in km-long fibers, where random birefringence fluctuations are present along the fiber length. We identify several distinct regimes that depend on the relative magnitude between the length scale of the random fluctuations and the beat-lengths of the interacting quasi-degenerate modes. In addition, we analyze the impact of polarization mode-dispersion and we demonstrate that random variations of the core radius, which are typically encountered during the drawing stage of the fiber, can represent the major source of bandwidth impairment. These results set a boundary on the limits of validity of the classical Manakov model and may be useful for the design of multimode parametric amplifiers and wavelength converters, as well as for the analysis of nonlinear impairments in long-haul spatial division multiplexed transmission.

\end{abstract}


%\IEEEpeerreviewmaketitle

\section{Introduction}

\IEEEPARstart{T}{he} last decade has been characterized by intense research in space-division multiplexing (SDM) schemes \cite{Richardson13} and novel all-optical devices for signal processing \cite{Wabnitz15}. Both these hot topics aim to develop new generation high-capacity internet networks capable of responding to the exponential growth of data demand. Within this framework, intermodal four-wave mixing (IM-FWM) in km-long multi-mode fibers (MMFs) is a key nonlinear process to be investigated for two main reasons. First, FWM is one of the main impairments affecting SDM transmissions \cite{Ellis13}. Second, the use of long fibers leads to large degrees of nonlinearity even at low input powers. This increases the overall efficiency of FWM-based devices and paves the way for the development of all-optical devices that may overcome the main limits associated with single-mode fiber-based devices. Specifically, the phase-matching condition in IM-FWM processes can be achieved far away from both the zero dispersion wavelength and the bandwidth of spontaneous Raman scattering, thus reducing the impact of the nonlinear cross-talk and of the Raman noise contribution \cite{Friis16}.

When analyzing light propagation in km-long fibers, it is important to take into account random birefringence fluctuations that occur on a length scale ranging from a few meters to several tens of meters \cite{Wuilpart01,Galtarossa04}. These fluctuations are caused by manufacturing imperfections, environmental variations or local stress mechanisms and impair the FWM dynamics by inducing linear coupling among quasi-degenerate modes. Recently, the experimental demonstration of IM-FWM in km-long fibers has been reported \cite{Friis16,Essiambre13}. However, while several theoretical works have addressed this issue in single-mode fibers \cite{McKinstrie04,Karlsson98,Guasoni12}, there are currently very few theoretical studies for MMFs \cite{Xiao14}.

In this paper we aim to provide a complete overview of the impact of random perturbations on the IM-FWM dynamics. We address several issues whose understanding will give useful guidelines for both  the mitigation of FWM for SDM transmission and for the design of all-optical devices for signal processing. The paper is organized as follows. In Section \ref{sec:II} we introduce the characteristic lengths that define the main fiber features in the presence of random perturbations. In Section \ref{sec:III} we model the random birefringence fluctuations and the induced linear coupling among quasi-degenerate modes of a MMF. In Section \ref{sec:IV} we highlight the existence of different FWM regimes related to the aforementioned characteristic lengths. Depending on their relative magnitude, the fiber may exhibit an "isotropic"-like behavior or a fully random coupling dynamics between quasi-degenerate modes which is described by the Generalized Manakov Model for MMFs \cite{Mumtaz13}. These results  set a boundary on the limits of validity of the classical Manakov model and therefore provide new perspectives for multimode long-haul transmission, similarly to what has recently been observed in single-mode fibers \cite{Marin17}. In Section \ref{sec:V} the impact of polarization mode dispersion on IM-FWM is discussed. Finally, in Section \ref{sec:VI}, we provide evidence that random fiber perturbations lead not only to a coupling between quasi-degenerate modes but also to fluctuations of the dispersion parameters of the different propagating spatial modes which can strongly impair the IM-FWM dynamics.

\section{Modeling of random birefringence fluctuations}
\label{sec:II}

Circular core isotropic fibers are characterized by groups of modes that are two-fold (groups $LP_{0m}$) or four-fold (groups $LP_{nm}$, with $n>1$ integer) degenerate. Degenerate modes of the same group possess identical dispersion parameters. However, in real optical fibers various random imperfections break the circular symmetry and isotropy of the fiber. Each degenerate mode is affected in a different way by these perturbations: as a result degenerate modes of the same group separate into a set of distinct quasi-degenerate modes, each one characterized by its own dispersive properties.

\begin{figure}
 	\includegraphics[width=1\columnwidth]{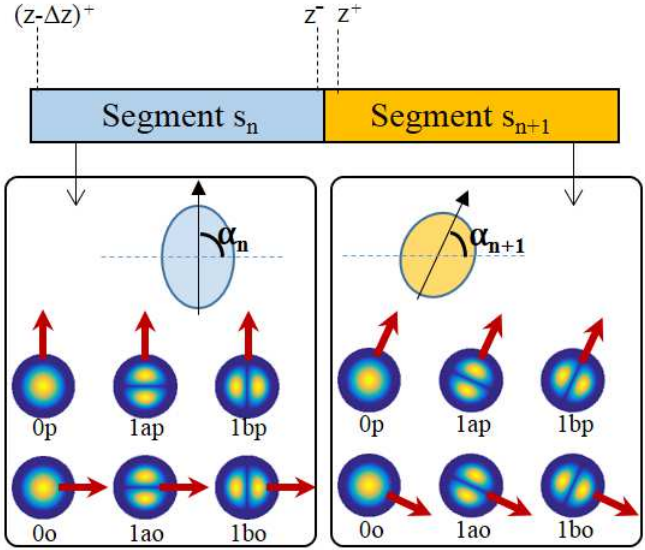}
 	\caption {Schematic of multimode randomly birefringent fiber. Two consecutive segments of fiber, $s_n$ ans $s_{n+1}$, are displayed. Each segment has its own cross-section shape  (elliptic in this figure)  characterized by a particular angle, $\alpha_n$ and $\alpha_{n+1}$ respectively, defining the direction of the fast birefringence axis (black solid arrow). Modes of groups $LP_{01}$ and $LP_{11}$ are shown for the two segments. Both the electric field (solid red arrow)  and the axis of symmetry of each mode are aligned parallel or orthogonal to the fast birefringence axis. Points $z^{-}$ and $z^{+}$ are also shown, representing respectively the positions immediately before and after the entry in segment $s_{n+1}$. Similarly, point $(z-\Delta z)^{+}$ is the position immediately after the entry in segment $s_n$.}
 	\label{FibreSchematic}
\end{figure}

The exact modeling of each source of perturbation is cumbersome and is still an active topic of research \cite{Palmieri04}. On the other hand, their global effect is that of a local and asymmetric weak variation of the fiber cross-section shape and size, as well as of the refractive index, giving rise to a weak local birefringence whose axes move randomly along the fiber \cite{Wai97}. We take as a reference the fast axis of the fiber and indicate with $\alpha(z)$ its angular orientation at the position $z$ along the fiber. As the perturbations are typically weak, we can safely assume that the shape of the modes is preserved. What changes instead is the orientation of their electric field, which is aligned to the local axes of birefringence and is thus either parallel or orthogonal to $\alpha(z)$ (see Fig.~\ref{FibreSchematic}). Furthermore, each mode has its own propagation constant $\beta(z)$, inverse group velocity $\beta_1(z)=\partial\beta/\partial\omega$ and chromatic dispersion $\beta_2(z)=\partial^2\beta/\partial\omega^2$ that are generally $z-$dependent.

In the following, for the sake of simplicity, we neglect higher-order dispersion and we assume the two groups of modes involved in the IM-FWM process are the $LP_{01}$ and the $LP_{11}$. Note however that the main outcomes in this paper can be easily generalized to include the interaction between different groups of modes and the presence of higher-order dispersion terms. We denote by $0p$ and $0o$ the two quasi-degenerate modes of group $LP_{01}$ that are respectively parallel (p) or orthogonal (o) to $\alpha$. Similarly, we denote by $1ap$, $1bp$, $1ao$ and $1bo$ the four quasi-degenerate modes of the group $LP_{11}$, as illustrated in Fig.~\ref{FibreSchematic}.

The angle $\alpha(z)$ changes randomly along the fiber length and is characterized by a correlation length, $L_C$, which defines the length-scale over which random perturbations become uncorrelated. As previously outlined, in typical standard fibers, $L_C$ varies from a few meters to some tens of meters.

For each pair of quasi-degenerate modes $m$ and $n$ we can define a corresponding beat-length $L_{B(m-n)}=2\pi/(\beta_m-\beta_n)$. In the problem under analysis there are 4 independent beat-lengths, $L_{B(0p-0o)}$, $L_{B(1ap-1ao)}$, $L_{B(1ap-1bp)}$ and $L_{B(1ap-1bo)}$, from which the 3 remaining beat-lengths can be computed (e.g., $L_{B(1bp-1ao)}^{-1}= L_{B(1ap-1ao)}^{-1}-L_{B(1ap-1bp)}^{-1} $). While beat-lengths are generally z-dependent, the FWM dynamics is mainly sensitive to their spatial average (see Section \ref{sec:IV}). Therefore, in what follows we refer to their spatial average.  The beat-length is indicative of the length scale over which the two quasi-degenerate modes acquire a significant phase-difference and thus of the minimum fiber length $L$ which is necessary to distinguish each other. Typically, the stronger the local perturbations, the larger the difference between the two propagation constants is, thus the shorter the corresponding beat-length. Therefore, the correlation length $L_C$ is a measure of "how fast" random perturbations occur, whereas the beat-lengths among quasi-degenerate modes measure "how strong" these perturbations are. If the fiber length $L$ is much shorter than all the beat-lengths, that is $L \ll \min\{|L_B|\}$, then modes within the same group propagate together in phase. In other words: in this instance random perturbations are weak enough so that the fiber can be considered perfectly circular and isotropic along its whole length. It is worth noting that beat-lengths can vary across a range of values from a few meters to tens of meters. For this reason, typical isotropic fibers are a few tens of meters long, so that relevant degrees of nonlinearity can only be achieved at the expenses of a large amount of input power. In the following, however, we are interested in km-long fibers, for which even small power levels may give rise to significant nonlinear effects. Therefore, in the following we can safely assume $L \gg \max\{|L_B|\}$, where $\max\{|L_B|\}$ is the largest beat-length.

The relative magnitude between the characteristic lengths discussed here gives rise to different FWM regimes that will be analyzed in the next sections.

\section{Random coupling induced by perturbations}
\label{sec:III}

To understand the coupling mechanism induced by random perturbations, it is useful to represent the fiber as a concatenation of  short segments of length $\Delta z$ (see Fig.~\ref{FibreSchematic}). Each segment is short enough to preserve, along its whole length, both the direction $\alpha$ of the birefringence axes and the dispersion parameters of all modes. Let us consider light propagation in two consecutive segments $s_n$ and $s_{n+1}$. Segment $s_n$ ($s_{n+1}$) is characterized by its own direction $\alpha_n$ ($\alpha_{n+1}$), with respect to which the electric field of the modes is parallel or orthogonal. We indicate with ${\bf A}(z)=[A_{0p}(z),A_{0o}(z),A_{1ap}(z),A_{1ao}(z),A_{1bp}(z),A_{1bo}(z)]$ the vector of the corresponding 6 modal amplitudes. Modes of segment $s_n$, immediately before entering $s_{n+1}$, are projected onto the modes of $s_{n+1}$. The projection is described by the following linear relation: ${\bf A_{n+1}^{(in)}} = {\bf P}{\bf A_n^{(out)}}$, where ${\bf A_n^{(out)}}\equiv {\bf A}(z^-)$ is the vector of amplitudes at point $z^-$, at the end of $s_n$ and just before entering $s_{n+1}$, and ${\bf A_{n+1}^{(in)}}\equiv {\bf A}(z^+ )$ is the vector at point $z^+$, just after entering $s_{n+1}$ (Fig.~\ref{FibreSchematic}). The  projection matrix reads:
\begin{equation}
{\bf P}=\left[ \begin{array}{cccccc}
C& -S & 0 & 0 & 0 & 0\\
S &  C & 0 & 0 & 0 & 0\\
0 & 0 & C & -S & -S & 0  \\
0 & 0 & S&  C &  0 & -S \\
0 & 0 & S & 0 & C & -S \\
0 & 0 & 0 &  S &  S & C
\label{P}
\end{array} \right]
\end{equation}
where $C=\cos(\Delta\alpha)$ and $S=\sin(\Delta\alpha)$, with $\Delta\alpha = \alpha_{n+1}-\alpha_{n}$. According to this model, coupling among different quasi-degenerate modes is thus induced by the random variation $\Delta\alpha$ of the birefringence axes. If no variation occurs, i.e. $\Delta\alpha=0$, then there is no energy exchange within quasi-degenerate modes (${\bf A_{n+1}^{(in)}}={\bf A_n^{(out)}}$), which is consistent with the assumption that the modal shape is largely preserved along the fiber length. Note also that according to matrix ${\bf P}$  there is no coupling between a mode of group $LP_{01}$ and a mode of group $LP_{11}$, which is typically the case in real fibers due to the large difference in their propagation constants.

In order to describe the propagation in the fiber, we study the evolution of light from point $(z-\Delta z)^+$, at the entry of segment $s_n$, to the point $z^+$, at the entry of $s_{n+1}$. First, light propagates through the segment $s_n$ from $(z-\Delta z)^+$ to $z^-$, undergoing both dispersion (operator $\hat{D}$) and nonlinearity (operator $\hat{N}$): ${\bf A_n^{(out)}} - {\bf A_n^{(in)}} = \Delta z[\hat{D}\{  {\bf A_n^{(in)}}  \} + \hat{N}\{  {\bf A_n^{(in)}}  \}] $, where ${\bf A_n^{(out)}} - {\bf A_n^{(in)}}$ indicates the mode amplitude variation, with ${\bf A_n^{(in)}}\equiv {\bf A}((z-\Delta z)^+)$. Then, modes of segment $s_n$ are projected onto modes of $s_{n+1}$ according to the relation ${\bf A_{n+1}^{(in)}} = {\bf P}{\bf A_n^{(out)}}$. From the two aforementioned relations, we finally evaluate the derivative $\partial {\bf A}/\partial z = \lim_{\Delta z\rightarrow 0} ({\bf A_{n+1}^{(in)}} - {\bf A_{n}^{(in)}})/\Delta z $, which after some algebra takes the form of the following Nonlinear Schr\"{o}dinger Equation (NLSE):

\begin{eqnarray}
\partial_z{\bf A} = &{\bf \bar{Q}}{\bf A} + (\bm{ \bar{\beta_1}}-v_r^{-1}){\partial_t\bf A} + i(1/2)\bm{\bar{\beta_2}}{\partial_{tt}\bf A}+\nonumber\\
& + \hat{N}\{  {\bf A} \}+\bm{ \tilde{\beta}}{\bf A} + \bm{ \tilde{\beta_1}}{\partial_t\bf A} + i(1/2)\bm{ \tilde{\beta_2}}{\partial_{tt}\bf A}
\label{OurNLSE}
\end{eqnarray}

In Eq.~(\ref{OurNLSE}) each dispersion coefficient $x=\bar{x}+\tilde{x}$ is separated into the sum of its spatial average $\bar{x}$ and its $z-$varying part $\tilde{x}$. This separation is introduced because, as will be discussed later, the averages and varying parts play a different role in the IM-FWM dynamics.

The 6x6 matrix ${\bf \bar{Q}}=2\pi{\bf \bar{L}_B}^{-1} + \partial_z\alpha{\bf U}$, where ${\bf \bar{L}_B} = diag[0, \bar{L}_{B(0p-0o)},0,\bar{L}_{B(1ap-1ao)}, \bar{L}_{B(1ap-1bp)},\bar{L}_{B(1ap-1bo)} ]$ and ${\bf U}$ is identical to $\bm{P}$, Eq.~(\ref{P}), for $\Delta\alpha=\pi/2$. Matrix $\bm{\bar{\beta_1}}=diag[\bar{\beta}_{1,0p},\bar{\beta}_{1,0o},\bar{\beta}_{1,1ap},\bar{\beta}_{1,1ao},\bar{\beta}_{1,1bp},\bar{\beta}_{1,1bo}]$ includes the average inverse group velocities, whereas $v_{r}$ is a free parameter that represents the velocity of a reference frame and can be conveniently chosen as $v_{r}=1/\bar{\beta}_{1,0p}$. Matrix $\bm{\bar{\beta_2}}=diag[\bar{\beta}_{2,0p},\bar{\beta}_{2,0o},\bar{\beta}_{2,1ap},\bar{\beta}_{2,1ao},\bar{\beta}_{2,1bp},\bar{\beta}_{2,1bo}]$ includes the average  chromatic dispersion coefficients. The  matrices $\bm{ \tilde{\beta_1}}$ and $\bm{ \tilde{\beta_2}}$ are formed analogously to $\bm{ \bar{\beta_1}}$ and $\bm{ \bar{\beta_2}}$ by replacing the average parameter $\bar{x}$ with the $z$-varying part $\tilde{x}$. Finally, matrix $\bm{\tilde{\beta}}=diag[\tilde{\beta}_{0p},\tilde{\beta}_{0o},\tilde{\beta}_{1ap},\tilde{\beta}_{1ao},\tilde{\beta}_{1bp},\tilde{\beta}_{1bo}]$. The operator $\hat{N}$ accounts for all nonlinear Kerr and Raman multimode interactions, as discussed in \cite{Poletti08}. In the following we assume that the chromatic dispersion of modes within the same group is the same, indicating with $\beta_ {2,0}$ the coefficients $\beta_ {2,0p}=\beta_ {2,0o}$ and with $\beta_ {2,1}$ the coefficients $\beta_ {2,1ap}=\beta_ {2,1a0}=\beta_ {2,1bp}=\beta_ {2,1bo}$.

It is worth noting that Eq.~(\ref{OurNLSE}) represents a generalization of the approach introduced in \cite{Wai97} to describe the effects of birefringence fluctuations in single-mode fibers. Note also that vector ${\bf A}$ in Eq.~(\ref{OurNLSE}) describes the modal amplitudes in the local reference frame, which is defined by the orientation $\alpha(z)$.

Similarly to previous work \cite{Xiao14}, in Eq.~(\ref{OurNLSE}) the overall effect of linear coupling is described by a 6x6 matrix ${\bf \bar{Q}}$. However, a major advantage of our approach is that this matrix is explicitly written in terms of the main real fiber parameters, that are the average beat-lengths among quasi-degenerate modes and the function $\alpha(z)$ which accounts for the random evolution of the birefringence axes. We can therefore study light propagation versus different profiles of $\alpha(z)$ and of beat-lengths, and identify different regimes that are discussed in the next Section.

\section{From Uncoupled to the Manakov regime}
\label{sec:IV}

\begin{figure}
 	\includegraphics[width=1\columnwidth]{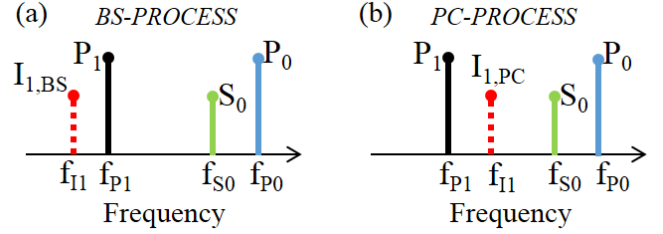}
 	\caption {(a) Representation of Bragg scattering. The input waves $P_0$, $P_1$ and $S_0$ generate the idler component $I_{1,BS}$. Energy conservation implies $f_{I1}-f_{P1}=f_{S0}-f_{P0}$. (b) Representation of phase conjugation. Energy conservation implies $f_{I1}-f_{P1}=f_{P0}-f_{S0}$.}
 	\label{ProcessSchematic}
\end{figure}

In this Section we study the impact of random perturbations on two important IM-FWM processes, namely Bragg scattering (BS) and phase conjugation (PC) \cite{McKinstrie02}. The configuration of the corresponding processes is represented in Fig.~\ref{ProcessSchematic}, where two input pumps $P_0$ and $P_1$ are coupled to modes $LP_{01}$ and $LP_{11}$, respectively, and an input seed signal $S_0$ is coupled to mode $LP_{01}$. Due to IM-FWM new idlers in the corresponding LP11 mode group are generated for both the BS ($I_{1,BS}$) and the PC processes ($I_{1,PC}$). All the waves are monochromatic and we indicate with $f_{P0}$, $f_{S0}$, $f_{P1}$ and $f_{I1}$ their corresponding frequency. 
We analyze the idler growth as a function of the system parameters, computing the idler power as the sum of the powers in the 4 quasi-degenerate $LP_{11}$ modes.

We initially assume that the dispersion coefficients are constant along the fiber length (that is, $\bm{ \tilde{\beta}}=\bm{\tilde{\beta_1}}=\bm{\tilde{\beta_2}}=0$ in Eq.~(\ref{OurNLSE})); the effects of their $z-$dependence will be discussed later. We also assume the pumps and signal are linearly copolarized, which maximizes the idler growth. For the fiber parameters used here we refer to the graded-index fiber employed in \cite{Friis16}. We fix $v_{0p}-v_{1ap}=100$ ps/km, whereas the chromatic dispersion coefficients are fixed respectively to $19.8$ ps/(nm km) for both modes of group $LP_{01}$ and to $21.1$ ps/(nm km) for all modes of group $LP_{11}$. The nonlinear overlap coefficients \cite{Poletti08} are indicated here as $C_{abcd}$. Indices $\{a,b,c,d\}$ are employed to refer to the modes: the index $0$ refers to one of the modes $0p$ and $0o$; $1$ refers to one of the modes $1ap$ and $1ao$ ; $2$ refers to one of the modes $1bp$ and $1bo$. Due to the symmetries of the modes under consideration, the coefficients with subscripts of the kind $aabb$ or $aaaa$ are the only non-zero ones and are invariant with respect to permutations of the indices (e.g. $C_{aabb}=C_{abab}$)  \cite{Friis16,Poletti08}. The nonlinear coefficients (related to the fiber discussed in \cite{Friis16}) are: $C_{0000}=0.63$~km$^{-1}$W$^{-1}$; $C_{0011}=C_{0022}=0.39$~km$^{-1}$W$^{-1}$; $C_{1111}=C_{2222}=0.60$~km$^{-1}$W$^{-1}$; $C_{1122}=0.18$~km$^{-1}$W$^{-1}$. Input powers are 22.5 dBm for each pump and 3.5 dBm for the signal. 

In order to get some realistic value for the strength of the random perturbations, we assume the average residual birefringence (that is the difference between the refractive indices of the birefringence axes) to be $\Delta n=1.5 \cdot 10^{-7}$, which is a typical value for standard optical fibers used in telecommunications. This provides an estimate for the beat-length $L_{B(0p-0o)} = \lambda/\Delta n = 10$ m and the inverse group velocity mismatch $\beta_{1,0p}-\beta_{1,0o}=\Delta n/c =0.5$ ps/km, where $\lambda=1550$ nm is the wavelength of the pump in mode $0p$. We use values of the same order for the beat-lengths and inverse group velocity mismatches of the group $LP_{11}$: $L_{B(1ap-1ao)}=25$ m; $L_{B(1ap-1bp)}=50$ m; $L_{B(1ap-1bo)}=8$ m; $\beta_{1,ap}-\beta_{1,1ao}= 0.2$ ps/km; $\beta_{1,ap}-\beta_{1,1bp}=0.4$ ps/km; $\beta_{1,1ap}-\beta_{1,1bo}=0.6$ ps/km.

The phase matching condition of IM-FWM processes in an isotropic fiber is fulfilled when the sum of the inverse group velocities of the pump and signal in group $LP_{01}$ equates to the sum of the inverse group velocities of the pump and signal in group $LP_{11}$ \cite{Friis16,Xiao14}. Therefore, phase matching is essentially related to the dispersion properties of the different mode groups. In randomly perturbed fibers, the small differences of group velocity among quasi-degenerate modes do not significantly affect the IM-FWM phase matching. Therefore the aforementioned phase-matching condition can be safely rewritten as $\beta_{1,0p}(f_{P0})+\beta_{1,0p}(f_{S0}) = \beta_{1,ap}(f_{P1})+\beta_{1,ap}(f_{I1})$.

\begin{figure}
 	\includegraphics[width=1\columnwidth]{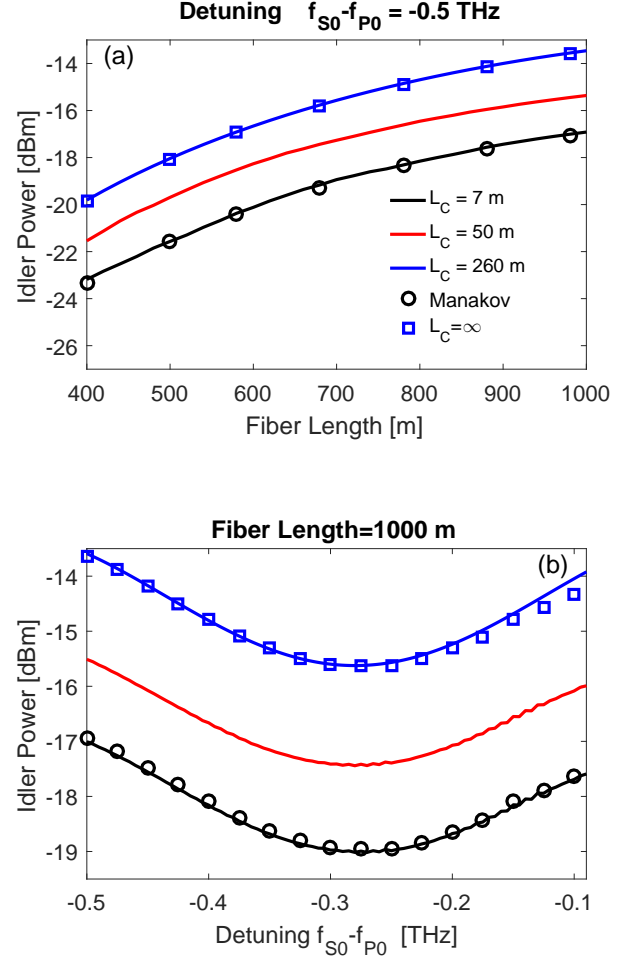}
 	\caption {BS process. (a) Idler power versus fiber length $L$ for a fixed signal-to-pump detuning $f_{S0}-f_{P0}=-0.5$ THz and for different values of coherence length $L_C$. The pump-to-pump detuning is $f_{P0}-f_{P1}=0.575$ THz, which maximizes the phase-matching. The case $L_C=\infty$ (fixed axes of birefringence) and the solution of the multimode Manakov model are also reported. (b) Idler power versus signal-to-pump detuning for a fixed fiber length $L=1000$ m.}
 	\label{BSResults}
\end{figure}

We first study the BS process (Fig.~\ref{ProcessSchematic}a). We simulate Eq.~(\ref{OurNLSE}) using the system parameters illustrated above and with a pump-to-pump detuning $f_{P0}-f_{P1}=0.575$ THz, which corresponds to the phase-matching condition for the BS process. The signal-to-pump detuning $f_{S0}-f_{P0}$ spans from $-0.5$ THz to $-0.1$ THz. Moreover, we generate random smooth profiles for $\alpha(z)$ with vanishing spatial average and different values of correlation length $L_C$. In our simulations $L_C$ is defined on the basis of the correlation function $C_{\alpha}(z)=|\int \alpha(z')\alpha(z'-z)dz'|/\int \alpha(z')^2dz'$; it indicates the length beyond which the correlation function remains below 0.1, that is $C_{\alpha}(z>L_C)<0.1$.
Simulation results are displayed in Fig.~\ref{BSResults} and show the existence of 3 distinct regimes depending on the relative magnitude between $L_C$ and the beat-lengths. For values of $L_C > 5\cdot \max\{|L_B|\}$, as in the case of $L_C=260$ m in Fig.~\ref{BSResults}, the idler dynamics does not depend on the particular value of $L_C$ and resembles the dynamics found for $L_C=\infty$, i.e. when the angle $\alpha(z)$ does not vary along the fiber length. In this instance, here named the \textit{Uncoupled Regime}, random perturbations evolve slowly enough to prevent any significant linear coupling among quasi-degenerate modes. Therefore, the fiber can be considered as a birefringent fiber with fixed axes of birefringence. As such, the idler growth strictly depends on the polarization direction of the  copolarized input waves and is maximized when they are aligned to one of the birefringence axes.

In the other extreme, when random fluctuations are fast and take place on a length scale shorter than the beat-lengths ($L_C < \min\{|L_B|\}$, as in the case of $L_C=7$ m in Fig.~\ref{BSResults}),  the idler growth computed by Eq.~(\ref{OurNLSE}) turns out to be in excellent agreement with the growth obtained by the solution of the multimode Manakov equations in the weak-coupling limit\cite{Mumtaz13}. Therefore we refer to this instance as the \textit{Manakov Regime} where, differently from the Uncoupled Regime, the system dynamics is independent of the polarization direction of the copolarized input beams. Following considerations similar to those discussed in \cite{Xiao14} we analytically estimate an idler amplification impairment of about -3.5 dB between the Uncoupled Regime and the Manakov Regime, which is  confirmed by our numerical results displayed in Fig.~\ref{BSResults}. Note that this impairment is almost independent of the signal frequency and the fiber length $L$. However, it is important to notice that this estimate applies only when, in the Uncoupled Regime, the input beams are aligned with one of the axes of birefringence, so that idler growth is maximized.

For intermediate values $\min\{|L_B|\} < L_C < 5\max\{|L_B|\}$  ($L_C= 50$ m in Fig.~\ref{BSResults}) we find an \textit{Intermediate Regime} where the idler amplification depends on the specific value of $L_C$.

\begin{figure}
 	\includegraphics[width=1\columnwidth]{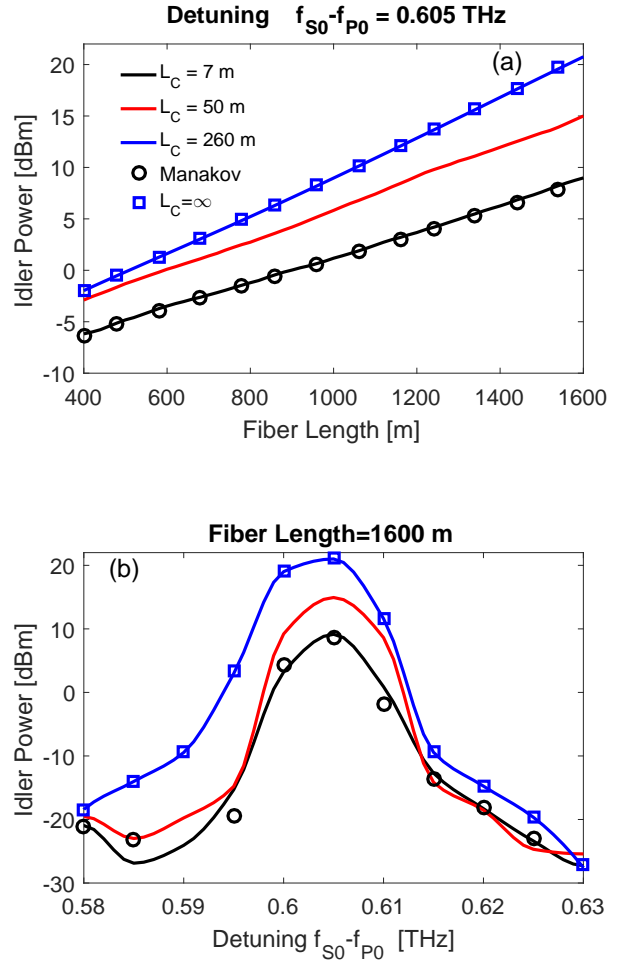}
 	\caption{PC process. (a) Idler power versus fiber length $L$ for different values of coherence length $L_C$. The signal-to-pump detuning $f_{S0}-f_{P0}=0.605$ THz maximizes the phase-matching. The pump-to-pump detuning is zero (degenerate FWM). The case $L_C=\infty$ (fixed axes of birefringence) and the solution of the multimode Manakov model are also reported. (b) Idler power versus signal-to-pump detuning for a fixed fiber length $L=1600$ m.}
 	\label{PCResults}
\end{figure}

Differently from the BS process, where phase matching is essentially governed by the pump-to-pump detuning, in the PC process (Fig.~\ref{ProcessSchematic}b) it is mainly related to the signal wavelength \cite{Friis16}. When both pumps are centered at the same frequency (degenerate FWM, $f_{P0}=f_{P1}$) we find that phase matching is optimized for $f_{S0}-f_{P0}=0.605$ THz.  Simulation results are displayed in Fig.~\ref{PCResults} when large  pump powers (32.5 dBm) are employed to get efficient idler amplification; the input signal power is -9 dBm. These results demonstrate once again the existence of the 3 distinct regimes observed in the BS process; on the other hand they also clearly highlight some particular differences with respect to the BS process, which are mainly related to the instability of the PC process. First, the idler power can significantly exceed the input signal power; and second, the idler amplification impairment induced by quick random  perturbations is not a constant value but is instead proportional to the fiber length as well as to the input pump powers. More precisely, we notice that for any value of $L_C$ the idler power (in dBm) versus fiber length is well approximated by a line with a slope that depends on $L_C$. Analytical considerations similar to those discussed in \cite{Xiao14} allow us to estimate an impairment of about $(2/3)C_{0011}(P_0P_1)^{1/2}$ between the slope in the Uncoupled Regime (when input waves are aligned to one of the birefringence axes) and the slope in the Manakov Regime.

We conclude this section by highlighting that in these simulations the beat-lengths have been chosen in such a way to achieve a strong linear coupling between the modes $LP_{11a}$ and $LP_{11b}$. However, as noticed in Ref. \cite{Xiao14}, the system dynamics is almost independent of their linear coupling. More specifically, in our study we find the same regimes and outcomes when the linear coupling between the modes $LP_{11a}$ and $LP_{11b}$ is null, which is one of the assumptions underlying the Manakov model \cite{Mumtaz13} . Consequently, our results indicate that the Manakov model can correctly describe the full FWM dynamics only when $L_C$ is of the same order as the shortest beat-length. On the other hand, even for realistic values of a few tens of meters (see e.g. $L_C=50$ m in Figs. \ref{BSResults}, \ref{PCResults}) the Manakov model fails in describing the idler dynamics, which sets important boundaries on its limits of applicability.

%We conclude this section by highlighting that these results indicate that the Manakov model can correctly describe the full FWM dynamics only when $L_C$ is of the same order as the shortest beat-length. On the other hand, even for realistic values of a few tens of meters (see e.g. $L_C=50$ m in Figs. \ref{BSResults}, \ref{PCResults}) the Manakov model fails in describing the idler dynamics, which sets important boundaries on its limits of applicability.
 
\section{Impact of polarization mode dispersion}
\label{sec:V}

So far we have neglected polarization mode dispersion (PMD), i.e., that the random dynamics discussed in the previous sections is in reality frequency dependent, similar to the case of single-mode fibers.

The main issue related to PMD is that waves at two different frequencies undergo a different randomization: as a result, the relative state of polarization (SOP) of  their spatial modes cannot be indefinitely preserved along the fiber. In the following analysis, the notion of spatial modes is related to the spatial shape of the transverse mode, therefore the group $LP_{11}$ is split into two distinct spatial modes $LP_{11a}$ and $LP_{11b}$.

The diffusion length $L_D=3/(4\pi^2D_p^2\Delta f^2)$ indicates the length scale beyond which the relative SOP is not maintained, where $\Delta f$ is the frequency detuning between the two waves and $D_p$ the PMD coefficient that is related to the beat-length through the relation $D_p=(2L_e)^{1/2}/(L_Bf)$, where $f$ is the carrier frequency of one of the two waves and $L_e$ the polarization correlation length \cite{AgrawalBook}. The latter is related to $L_C$ and $L_B$ according to a relation that varies depending on the regime under analysis \cite{Wai97}. Note that in the multimode system considered here a PMD coefficient and the related diffusion length can be defined for each one of the beat-lengths. Here we are mainly interested in the smallest, $\min\{L_D\}$, and largest, $\max\{L_D\}$, diffusion lengths. These two lengths allow us to distinguish two regimes: a low-PMD regime, when the fiber length $L \ll \min\{L_D\}$, where the relative SOP of spatial modes is preserved along the fiber;  and a high-PMD regime, when $L \gg \max\{L_D\}$, where the relative SOP varies randomly along the fiber.

\begin{figure}
 	\includegraphics[width=1\columnwidth]{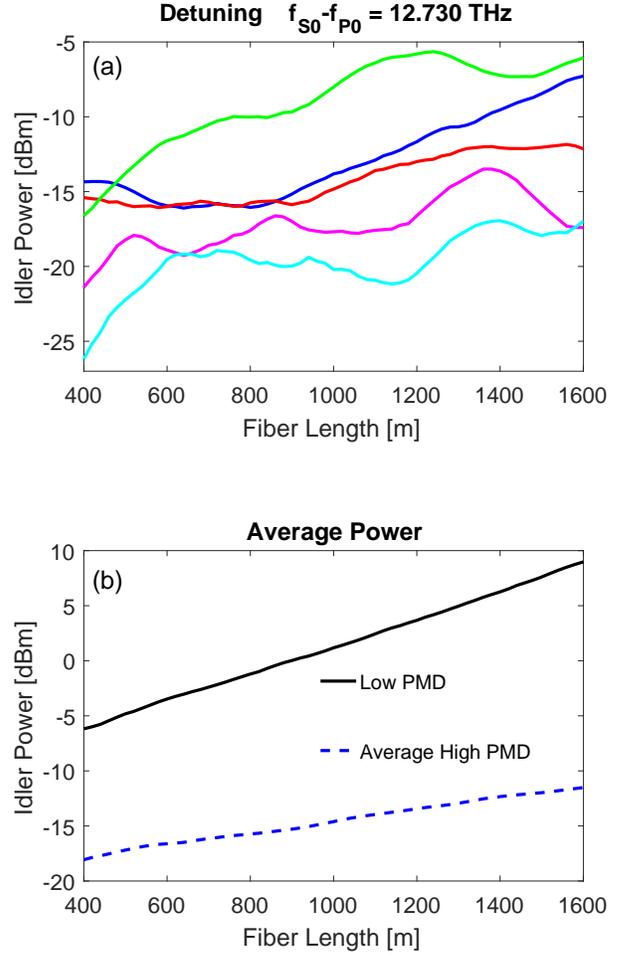}
 	\caption {(a) Idler power versus fiber length $L$ of the PC process in the high-PMD regime. Five different curves are displayed, each one corresponding to a different random realization of the function $\alpha(z)$ for the same correlation length $L_C=7$ m. System parameters are the same as for results displayed in Fig.~\ref{PCResults}(a), except for the GVD values of modes that have been modified to shift the phase-matching detuning $f_{S0}-f_{P0}$ to $12.730$ THz. (b) Comparison between the idler power in the low-PMD regime (phase matching detuning $f_{S0}-f_{P0}=0.605$ THz, see also Fig.~\ref{PCResults}) and the power averaged over 60 realizations in the high-PMD regime (phase matching detuning $f_{S0}-f_{P0}=12.730$ THz). }
 	\label{PMDResults}
 \end{figure}

An exhaustive analysis of the PMD impact is complex and out of the scope of the current paper, therefore in the following we limit our study to the Manakov Regime, which is important for km-long fibers, and we focus on the degenerate PC process introduced in Section \ref{sec:IV}. According to the system parameters, when $L_C=7$ m (Manakov Regime), the minimum diffusion length is several tens of km, therefore results displayed in Figs.\ \ref{BSResults}, \ref{PCResults} concern the low-PMD regime. In order to move towards the high-PMD regime of the PC process we keep the system parameters unchanged except for the chromatic dispersion coefficients that are reduced to $0.78$ ps/(nm km) for modes of group $LP_{01}$ and to 1.18 ps/(nm km) for modes of group $LP_{11}$. Consequently, the phase matching detuning $\Delta f=f_{S0}-f_{PO}$ increases to $12.73$ THz and the corresponding largest diffusion length $\max\{L_D\}$ decreases to $450$ m. In Fig.~\ref{PMDResults}(a) the idler power versus fiber length is displayed for different realizations of the orientation angle $\alpha(z)$ that are \textit{all} characterized by the same correlation length $L_C=7$ m. Differently from the low-PMD regime, where the idler growth is almost independent of the particular realization, the high-PMD regime is characterized by severe variations of the idler amplification from one realization to another. Note that a similar dynamics has been previously observed  in single-mode fibers \cite{Lin04}.Furthermore, from Fig.~\ref{PMDResults}(b) we notice that the idler growth, averaged over a consistent number of different realizations, is generally reduced by several dB due to PMD. The impairment is proportional to the fiber length and can be as large as several tens of dB at the output of a km-long fiber. These results clearly indicate that PMD puts a strong limit on the maximum bandwidth of multimode parametric devices based on km-long fibers.

\section{Impact of the variation of dispersive parameters}
\label{sec:VI}

The study of random perturbations portrayed in previous sections was based on the assumption that the dispersion parameters were constant along the fiber length ($\bm{ \tilde{\beta}}=\bm{\tilde{\beta_1}}=\bm{\tilde{\beta_2}}=0$ in Eq.~(\ref{OurNLSE})). More realistically however, random local perturbations affect these dispersion parameters and we will study the impact of this z-dependence in the following.

Towards this, we distinguish the \textit{intragroup dispersive parameters}, which are the beat-lengths ($L_{B(0p-0o)}$, $L_{B(1ap-1ao)}$, $L_{B(1ap-1bp)}$, $L_{B(1ap-1bo)}$) and the relative inverse group velocities ($\beta_{1,0p}-\beta_{1,0o}$, $\beta_{1,ap}-\beta_{1,ao}$, $\beta_{1,ap}-\beta_{1,bp}$, $\beta_{1,ap}-\beta_{1,bo}$) among quasi-degenerate modes of the same group, and  the \textit{intergroup dispersive parameters} $\beta_{1,0p}$, $\beta_{1,ap}$, $\beta_{2,0}$ and $\beta_{2,1}$, which describe the different dispersive properties of different modal groups.

As pointed out in Section \ref{sec:IV}, the IM-FWM dynamics at  phase-matching depends on the intragroup parameters, more precisely the beat-lengths, so that three distinct regimes can be distinguished which we called the Uncoupled, Manakov, and Intermediate Regimes. On the other hand, the phase matching condition for the BS and PC nonlinear processes between LP01 and LP11 modes depends on the intergroup parameters but is practically unaffected by intragroup parameters. Therefore, we proceed by studying separately the effect of the z-dependence of the intergroup and intragroup parameters.

Initially,  Eq.~(\ref{OurNLSE}) is solved by keeping the intergroup parameters fixed while varying  the intragroup parameters $\bm{\tilde{\beta}}$ and $\bm{\tilde{\beta_1}}$ with $z$. In order to implement the z-dependence, each intragroup parameter is defined as a random function $p(z)$ with spatial average $\bar{p}$, standard deviation $\sigma(p)$ and correlation length $L_C$.  Our numerical simulations show that the IM-FWM dynamics is not sensitive to local variations of the intragroup parameters, but only to their average value. We verified numerically that this is true even for standard deviations which are as large as the average value. It is worth noting that this outcome is in line with previous studies in single-mode fibers \cite{Wai97}. Therefore, we still recognize the three regimes found in Section \ref{sec:IV}, provided that the values of the beat-lengths are replaced by their spatial averages, so that the thresholds for the Uncoupled and Manakov regimes become $L_C > 5\max\{{\bar L_B}\}$ and $L_C < \min\{{\bar L_B}\}$, respectively.

Contrary to the case of intragroup parameters, even small variations of the intergroup parameters can strongly impact the phase-matching condition and then severely affect the IM-FWM dynamics. This issue has already been addressed in single-mode fibers, where small fluctuations of the  chromatic dispersion along the fiber can lead to a remarkable reduction of the idler amplification bandwidth \cite{Karlsson98,Yaman04}. In the multimode dynamics this issue is even more critical, as  the idler amplification band depends on all four intergroup parameters, that is, not only on the chromatic dispersion coefficients but also on the group velocities of groups $LP_{01}$ and $LP_{11}$ \cite{Friis16}.

As a practical example, we consider here a 1-km long MM silica step-index fiber whose core radius $R(z)$ varies randomly in $z$, thereby inducing fluctuations of the intergroup parameters. Note that radius fluctuation is a typical perturbation occurring on a length-scale $L_C$ of a few meters during the drawing stage of the fiber, where the standard deviation of the radius can be as large as $1\%$ \cite{Karlsson98}. Here we assume the core radius $R(z)$ to have average $\bar{R}=40$ $\mu$m and correlation length $L_C=5$ m. We simulate two distinct instances where its standard deviation is either $0.5\%$  ($\sigma(R)=2$ $\mu$m) or $1\%$  ($\sigma(R)=4$ $\mu$m), respectively.  The core-cladding index difference is fixed at 0.0025, independently of the wavelength. Several groups of modes can propagate in this fiber, but we assume only modes $0p$ and $1ap$ are excited and propagate. Note that in order to isolate the impact of the variation of intergroup parameters here we do not introduce random linear coupling among quasi-degenerate modes (i.e. we set $\partial_z\alpha=0$ in Eq.~(\ref{OurNLSE})).
We calculate the intergroup parameters and nonlinear coefficients at each position $z$ of the fiber, and solve Eq.~(\ref{OurNLSE}) accordingly. In order to compute the intergroup parameters at position $z$, we first derive the propagation constants $\beta_{0p}(z)$ and $\beta_{1ap}(z)$ of modes $0p$ and $1ap$ by solving the modal characteristic equation for a circular-core fiber of radius $R(z)$; we then directly infer the intergroup parameters $\beta_{1,0p}(z)=\partial \beta_{0p}/\partial\omega$, $\beta_{1,1ap}(z)=\partial \beta_{1ap}/\partial\omega$, $\beta_{2,0}=\partial^2 \beta_{0p}/\partial\omega^2$ and $\beta_{2,1}=\partial^2 \beta_{1ap}/\partial\omega^2$ by also taking into account the material dispersion of silica. Similarly, in order to compute the nonlinear coefficients we first calculate the transverse mode profiles as a function of $R(z)$ and then the nonlinear overlap integrals. In this way, our numerical simulations also account for the z-dependence of the nonlinear coefficients. We do this for completeness, however we anticipate that these fluctuations of the nonlinearity have very little effect on the IM-FWM dynamics, so that in practice their average value could safely be used in simulations.

\begin{figure}
 	\includegraphics[width=1\columnwidth]{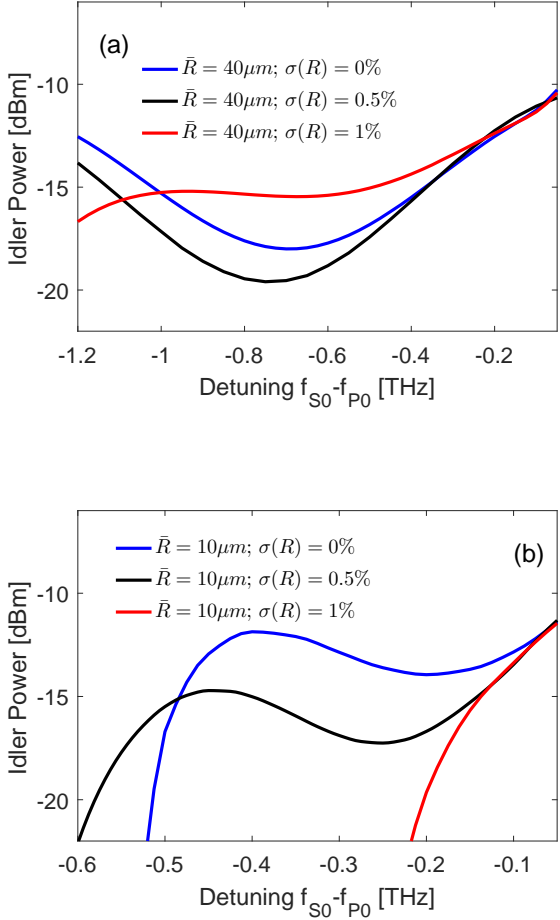}
 	\caption {Idler power versus signal-to-pump detuning for the BS process at the fiber output. The cases of fixed radius and z-varying radius ($\sigma(R)$=0.5$\%$ and $\sigma(R)$=1$\%$) are displayed. In (a) the average radius $\bar{R}=40$ $\mu$m; in (b) $\bar{R}=10$ $\mu$m.}
 	\label{VaryingRadiusResults}
 \end{figure}

When solving Eq.~(\ref{OurNLSE}) for the BS process, we fix the pump-to-pump detuning to $f_{P0}-f_{P1}=1.65$ THz, which corresponds to the phase matching condition for a fiber with constant radius $R=\bar{R}=40$ $\mu$m. In Fig.~\ref{VaryingRadiusResults}(a) the output idler power is plotted versus the signal-to-pump detuning in both the cases of fixed and varying radius. We note that the idler amplification bandwidth is only slightly impaired by fluctuations of the intermodal parameters. When repeating the same analysis with a bimodal fiber of average radius ${\bar R}=10$ $\mu$m (phase-matching at $f_{P0}-f_{P1}=5.095$ THz) and standard deviation  $0.5\%$ or $1\%$ (that is $\sigma(R)=0.05$ $\mu$m or $\sigma(R)=0.1$ $\mu$m, see Fig.~\ref{VaryingRadiusResults}(b)), we find instead that the amplification bandwidth is severely reduced. This is explained by the fact that the smaller the radius, the more the modes spread out in the outer core, such that their dispersion parameters become strongly sensitive to variations of the core size. Consequently, the phase matching condition cannot be preserved along the fiber length, which causes the drastic reduction of bandwidth. Note that almost the same results are found when introducing linear coupling among quasi-degenerate modes, except for an amplification impairment of about $-3.5$ dB, as pointed out in Section \ref{sec:IV}.

It is worth noting that in Ref. \cite{Friis16} the authors have studied the BS process in a 1 km-long bimodal fiber and found that the experimental bandwidth at -3 dB was about 4 times narrower than the bandwidth  estimated in numerical simulations when considering the propagation in a totally uniform fiber. They then conjectured that fluctuations of the dispersive parameters may be the principal source of the observed discrepancy. The plots in Fig.\ref{VaryingRadiusResults}(b), related to the bimodal fiber of average ${\bar R}=10$ $\mu$m, give support to this interpretation.

More in general, the results displayed in Fig.\ref{VaryingRadiusResults} demonstrate that the analysis of the device robustness against fluctuations of the relative intergroup dispersive parameters is an essential step when designing multimode parametric devices in km-long fibers. It is worth noting from Fig.~\ref{VaryingRadiusResults}(b) that these fluctuations may completely suppress the idler growth even when the frequency detuning among waves is low (that is, in a low-PMD regime). Therefore they may constitute the dominant factor of bandwidth impairment in parametric amplifiers. On the other hand, the same effect may be an interesting tool to exploit in order to reduce FWM impairments in SDM transmissions.

\section{Conclusion}
\label{sec:VII}

In this paper we studied the IM-FWM dynamics taking place between different groups of modes in a km-long MMF. We identified three distinct regimes which depend on the relative magnitude between the correlation length $L_C$ of random longitudinal fiber fluctuations and the beat-lengths of the interacting quasi-degenerate modes. We then demonstrated that the Manakov model can reproduce the FWM dynamics only when $L_C$ is of the same order of or shorter than the smallest beat-length. On the contrary, when $L_C$ is much longer than all beat-lengths, the fiber acts as a birefringent fiber with fixed axes of birefringence where the IM-FWM dynamics strictly depends on the relative polarization of the input waves with respect to the axes of birefringence (Uncoupled Regime). The maximum amplification impairment between the Uncoupled and the Manakov regime varies depending on the kind of FWM process being considered: for BS processes it is about $-3.5$ dB almost independently of the fiber length and input pump powers, whereas for PC processes it is directly proportional to both the fiber length and the pump powers. PMD is a further source of impairment: in the high-PMD regime not only is the amplification substantially reduced, but it also depends on the particular longitudinal profile of the fiber perturbations. Therefore, two different profiles $\alpha(z)$ with the same correlation length may lead to strongly different FWM amplification. Finally, we highlighted that random fiber fluctuations induce a random evolution of the dispersion parameters of the different groups of modes. This can result in a severe reduction of the FWM bandwidth and thus constitutes one of the major issues when addressing the design of efficient multimode devices for parametric amplification/conversion.

Overall, these results shed light on the FWM dynamics in km-long MMFs, and as such could find useful application in the study of SDM transmission as well as in the development of multimode devices for all-optical signal processing, which are building-blocks of future all-optical networks. Finally, the finding of different FWM regimes that are not captured by the Manakov model raises important questions on its limits of validity and paves the way towards novel and robust transmission formats in multimode systems, similar to what was recently demonstrated in  single-mode fibers \cite{Marin17}.

\section*{Acknowledgments}
The data for this work is accessible through the University of Southampton Institutional Research Repository (DOI:xxxxxx)
%\appendices

%\section{}


% \end{IEEEbiography}


%\newpage



%\vfill




\begin{thebibliography}{99}

\bibitem{Richardson13}
D.~J. Richardson, J.~M. Fini and L.~E. Nelson, \emph{Space-division multiplexing in optical fibres}, Nature Photon. 7, 354-362 (2013). 

\bibitem{Wabnitz15}
S.~Wabnitz and B. J.~Eggleton, \emph{All-optical signal processing}, Springer (2015). 

\bibitem{Ellis13}
A. D. Ellis, N. Mac Suibhne, F. C. Garcia Gunning, and S. Sygletos, \emph{Expressions for the nonlinear transmission performance of multi-mode optical fiber}, Opt. Exp. 21, 22834-22846 (2013).



\bibitem{Friis16}
S. M. M. Friis, I. Begleris, Y. Jung, K. Rottwitt, P. Petropoulos, D. J. Richardson, P. Horak and F. Parmigiani, \emph{Inter-modal four-wave mixing study in a two-mode fiber}, Opt. Exp. 24, 30338-30349 (2016).


\bibitem{Wuilpart01}
M.~Wuilpart, P.~M\'egret, M.~Blondel, A. J.~Rogers and Y.~Defosse, \emph{Measurement of the spatial distribution of birefringence in optical fibers}, IEEE Photon. Technol. Lett. 13, 836-838 (2001).

\bibitem{Galtarossa04}
A.~Galtarossa and L.~Palmieri, \emph{Spatially resolved PMD measurements}, J. Lightw. Technol. 22, 1103-1115 (2004).

\bibitem{Essiambre13}
R.-J.~Essiambre, M. A.~Mestre, R.~Ryf, A. H.~Gnauck, R. W.~Tkach, A. R.~Chraplyvy, Y.~Sun, X.~Jiang, and R.~Lingle, \emph{Experimental investigation of inter-modal four-wave mixing in few-mode fibers}, IEEE Photon. Technol. Lett. 25, 539-542 (2013).

\bibitem{McKinstrie04}
C. J.~McKinstrie, H.~Kogelnik, R. M.~Jopson, S.~Radic and A. V.~Kanaev, \emph{Four-wave mixing in fibers with random birefringence}, Opt. Exp. 12, 2033-2055 (2004).

\bibitem{Karlsson98}
M.~Karlsson, \emph{Four-wave mixing in fibers with randomly varying zero-dispersion wavelength}, J. Opt. Soc. Am. B 15, 2269-2275 (1998).

\bibitem{Guasoni12}
M.~Guasoni, V. V.~Kozlov and S.~Wabnitz, \emph{Theory of polarization attraction in parametric amplifiers based on telecommunication fibers}, J. Opt. Soc. Am. B 29, 2710-2720 (2012).

\bibitem{Xiao14}
Y.~Xiao, R.-J.~Essiambre, M.~Desgroseilliers, A. M.~Tulino, R.~Ryf, S.~Mumtaz and G. P.~ Agrawal, \emph{Theory of intermodal four-wave mixing with random linear mode coupling in few-mode fibers}, Opt. Exp. 22, 32039-32059 (2014). 

\bibitem{Mumtaz13}
S.~Mumtaz, R.-J.~Essiambre, G. P.~Agrawal, \emph{Nonlinear propagation in multimode and multicore Fibers: generalization of the Manakov equations}, J. Lightw. Technol. 31, 398-406 (2013).

\bibitem{Marin17}
M.~Gilles, P. Y.~Bony, J.~Garnier, A.~Picozzi, M.~Guasoni and J.~Fatome, \emph{Polarization domain walls in optical fibres as topological bits for data transmission}, Nature Photon. 11, 102-107 (2017).


\bibitem{Palmieri04}
L.~Palmieri and A.~Galtarossa, \emph{Coupling effects among degenerate modes in multimode optical fibers}, IEEE Photon. J. 6, 0600408 (2004).


\bibitem{Wai97}
P. K. A. Wai and C. R. Menyuk, \emph{Polarization mode dispersion, decorrelation, and diffusion in optical fibers with randomly varying birefringence}, J. Lightw. Technol. 14, 148-157 (1997).

\bibitem{Poletti08}
F.~Poletti and P.~Horak, \emph{Description of ultrashort pulse propagation in multimode optical fibers}, J. Opt. Soc. Am. B 25, 1645-1654 (2008).

\bibitem{McKinstrie02}
C. J.~McKinstrie, S.~Radic, and A. R.~Chraplyvy, \emph{Parametric amplifiers driven by two pump waves},IEEE J. Sel. Top. Quantum Electron. 8, 538-547 (2002).

\bibitem{AgrawalBook}
G. P.~Agrawal, \emph{Nonlinear Fibre Optics, 4th Edition}, Elsevier (2007).

\bibitem{Lin04}
Q.~Lin and G. P.~Agrawal, \emph{Effects of polarization-mode dispersion on fiber-based parametric amplification and wavelength conversion}, Opt. Lett. 29, 1114-1116 (2004). 


\bibitem{Yaman04}
F.~Yaman, Q.~Lin, S.~Radic and G. P.~Agrawal, \emph{Impact of dispersion fluctuations on dual-pump fiber-optic parametric amplifiers}, IEEE Photon. Techol. Lett. 16, 1292-1294 (2004).



  %0.5em minus 0.4em\relax Harlow, England: Addison-Wesley, 1999.

\end{thebibliography}
\end{document}